\documentclass[prb,twocolumn,showpacs,superscriptaddress,preprintnumbers,amssymb]{revtex4}
\usepackage{graphicx}
\usepackage{color}
\usepackage{dcolumn}
\usepackage{bm}
\usepackage{wasysym}
\usepackage{graphicx}

\newcommand{\beq}{\begin{equation}}
\newcommand{\eeq}{\end{equation}}
\newcommand{\beqn}{\begin{eqnarray}}
\newcommand{\eeqn}{\end{eqnarray}}

\newcommand{\ua}{\uparrow}
\newcommand{\da}{\downarrow}
\newcommand{\ra}{\rightarrow}

\newcommand{\cL}{ {\cal L} }

\newcommand{\cT}{ {\cal T} }

\newcommand{\vect}[1]{{\bm{#1}}}

\newcommand{\ii}{\mathrm{i}}

\newcommand{\hV}{\hat{V}}

\newcommand{\SO}{\mathrm{SO}}

\newcommand{\SU}{\mathrm{SU}}
\newcommand{\U}{\mathrm{U}}

\begin{document}


\title{Superconductor-Insulator Transition in TMD moir\'{e} systems and the \\ Deconfined Quantum Critical Point }

\author{Nayan Myerson-Jain}

\author{Cenke Xu}

\affiliation{Department of Physics, University of California,
Santa Barbara, CA 93106}

\date{\today}

\begin{abstract}

We propose that the recently observed superconductor-insulator
transition (SIT) in a twisted bilayer transition metal
dichalcogenides moir\'{e} system~\cite{TMDSC} at hole filling $\nu
= 1$ may be described by the deconfined quantum critical point
(DQCP), which was originally proposed for the transition between
the N\'{e}el order and the valence bond solid (VBS) order on the
square lattice~\cite{deconfine1,deconfine2}. The key symmetries
involved in the original DQCP include a $\SO(3)_s$ spin symmetry,
as well as a $C_4$ lattice rotation symmetry for the VBS order
that is enlarged into a $\U(1)_v$ symmetry near the DQCP. In the
current SIT under consideration, the counterpart of the $\SO(3)_s$
spin symmetry is an approximate $\SO(3)_v$ symmetry that
transforms between different crystalline orders on the triangular
lattice, and the role of the $\U(1)_v$ symmetry is replaced by the
ordinary charge-$\U(1)_e$ symmetry. At the DQCP, the $\SO(3)_v
\times \U(1)_e$ may enlarge into an emergent SO(5) symmetry. Under
strain, the SIT is driven into either a prominent first order
transition, or an ``easy-plane" DQCP, which is expected to have an
emergent O(4) symmetry.

\end{abstract}

\maketitle


\section{Introduction}


The recent rapid experimental progress of searching for novel
correlated states of matter in moir\'{e} systems has generated
continuous excitement in the condensed matter community. Various
exotic phases of matter, including the fractional Chern
insulator~\cite{youngFCI,harvardFCI}, fractional quantum anomalous
Hall effects~\cite{xuFQAH1,xuFQAH2,makFCI,juFQAH,shFQAH}, and even
signatures of fractional topological insulators~\cite{FTIex} have
all been reported. Most recently, superconductivity has also been
reported in twisted bilayer transition metal dichalcogenides (TMD)
moir\'{e} systems at hole filling $\nu = 1$ (one hole per
moir\'{e} unit cell)~\cite{TMDSC,TMDSC2}, and a continuous
superconductor-insulator transition (SIT) tuned by the
displacement field was observed~\cite{TMDSC}. Given all the
remarkable unconventional phases identified in the TMD moir\'{e}
systems, it is natural to ask whether the quantum phase
transitions in these systems can also transcend the ordinary
paradigm. The goal of this work is to propose that the SIT
observed in Ref.~\onlinecite{TMDSC} may be described by the
deconfined quantum critical point
(DQCP)~\cite{deconfine1,deconfine2}, an archetypal example of
unconventional QCP beyond the standard paradigm.

The original DQCP was proposed as an unconventional quantum phase
transition between the N\'{e}el order with spontaneous symmetry
breaking (SSB) of the SO(3)$_s$ spin symmetry, and a valence bond
solid (VBS) order which spontaneously breaks the lattice $C_4$
rotation symmetry, and at the DQCP the $C_4$ symmetry is enlarged
into a $\U(1)_v$ emergent symmetry. The DQCP has been an extremely
active subfield in the theoretical and numerical condensed matter
community~\cite{senthilreview}, and it has very profound
connections to many other notions in theoretical physics, such as
the 't Hooft anomaly, emergent
symmetry~\cite{senthilfisher,JQ1,JQ2,loopmodel1,loopmodel2,Sun_2021},
the symmetry protected topological (SPT)
phases~\cite{wenspt,wenspt2} in one higher
dimension~\cite{senthilashvin}, the web of
duality~\cite{SO5,dualreview}, the ``loss of
conformality"~\cite{conformallost,SO5,hefuzzy}, and the
intrinsically gapless SPT phases~\cite{igspt1,igspt2}. But the
study on DQCP has been mostly theoretical and numerical, except
for a possible experimental realization of the DQCP with certain
anisotropy in
SrCu$_2$(BO$_3$)$_2$~\cite{dqcpex,dqcpex2,dqcpex3,dqcpex4,Leedqcpex},
which was expected to be connected to the ``easy-plane"
DQCP~\cite{maxO4}. But the experimental connection of the ``full"
DQCP with supposedly an emergent SO(5) symmetry has been elusive.
In this work we hope to bridge the gap of the experimental
connection of the DQCP.

The twisted bilayer TMD moir\'{e} system with hole filling can be
modelled as a two-orbital continuum model for each spin/valley
flavor, one orbital from each
layer~\cite{tmdTI,Yu_2019,Devakul_2021}. The tunnelling amplitude
as well as the energy difference between the two layers together
form a three component vector $\mathbf{\Delta}(\mathbf{r})$, which
depends on the spatial coordinate $\mathbf{r}$ due to the
twisting. The configuration of $\mathbf{\Delta}(\mathbf{r})$ may
have a nonzero integer Skyrmion number within the moir\'{e} unit
cell, which leads to a nonzero Chern number for the moir\'{e}
minibands for each spin/valley index. At small twisting angle, the
physics of the first two moir\'{e} minibands can be captured by
the Kane-Mele tight-binding model~\cite{kane2005a,kane2005b} on
the honeycomb moir\'{e} lattice. A displacement field in the
twisted bilayer TMD moir\'{e} system explicitly breaks the
exchange symmetry between the A/B sublattice (the $C_{2y}$
symmetry) of the moir\'{e} honeycomb lattice, i.e. the holes
prefer to stay on (for example) the sublattice A of the moir\'{e}
lattice, which forms a triangular lattice. With larger twisting
angle, at least one more sublattice at the center of each hexagon
is needed to reproduce the physics of the moir\'{e}
minibands~\cite{millisbridge}, but a displacement field still
lifts the degeneracy among the three sublattices. Although the
superconductivity happens when the holes are at half-filling of a
pair of Chern bands with opposite Chern numbers, the nontrivial
topology of the Chern bands may {\it not} play a crucial role at
the observed superconductor-insulator transition, as the
superconductor could still be ordinary rather than a topological
superconductor. In fact, once the charge $\U(1)_e$ is broken, a
quantum spin Hall insulator can be smoothly connected to an
ordinary superconductor.

If the holes are gapped on both sides of the SIT, one can just
focus on the bosonic sector of the system, if the goal is to
understand the universality class of the SIT. It was argued (for
example in Ref.~\onlinecite{Balents_2005}) that the universal low
energy features of certain SITs can be understood as
superfluid-insulator transitions of interacting bosons. Given the
observations in the last paragraph, and the ``strong pairing"
nature of the superconductor phase~\cite{TMDSC}, we may consider
the SIT as a superfluid-insulator transition of hard core bosons
on an effective triangular lattice at half-filling, i.e. on
average half-boson per site. The hard core boson carries the same
quantum number as a Cooper pair. The SIT can also be modelled as
the low energy bosonic sector of interacting spin-1/2 fermions on
the effective triangular lattice, also at half-filling. A direct
and continuous SIT in this case must be beyond the ordinary
Wilson-Fisher paradigm, at least in the limit with no disorder. In
fact half-filled bosons (or spin-1/2 fermions) on a triangular
lattice is prohibited from having a trivial insulator (gapped and
nondegenerate), due to the Lieb-Shultz-Mattis
theorem~\cite{LSM,hastings,oshikawa}. More specifically, the
insulator must either spontaneously break certain symmetry, or it
must form a gapless or topological spin liquid.
\begin{center}
\begin{figure}
\includegraphics[width=0.45\textwidth]{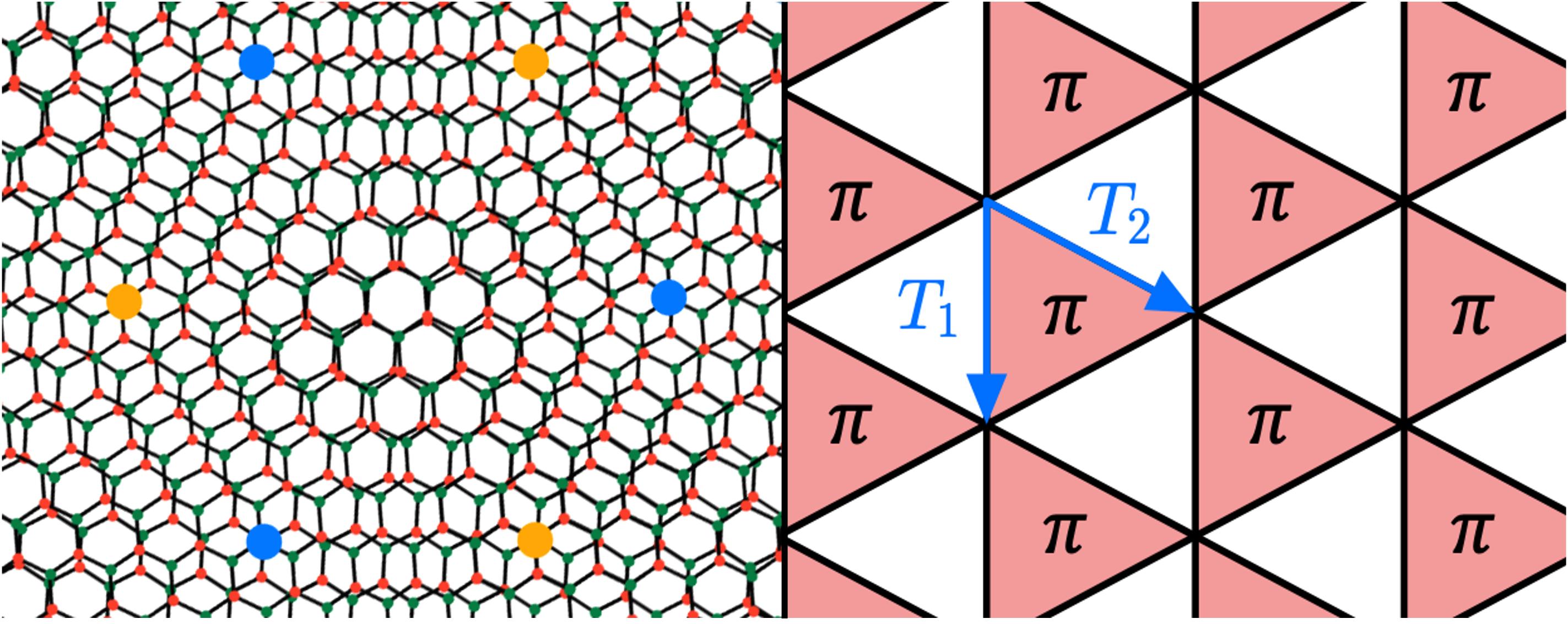}
\caption{The moir\'{e} lattice ($a$), and the effective triangular
lattice ($b$) for the study of the SIT. The mean field state of
the fermionic parton $\psi_\alpha$ sees a $\pi$-flux on half of
the triangle plaquettes ($b$).} \label{lattice}
\end{figure}
\end{center}

If the insulator has a $Z_2$ topological order, there is a
previously known mechanism for the SIT. In this scenario, the
insulator phase in the phase diagram is a $Z_2$ spin
liquid~\cite{subirz2,wenz2,wensl,sondhisl}. A $Z_2$ spin liquid on
the triangular lattice does not need to break any lattice
symmetry. The simplest $Z_2$ spin liquid can be constructed with a
parton formalism, $c_{j,\alpha} = b_j f_{j,\alpha}$:
$c_{j,\alpha}$ corresponds to the physical fermion ($e.g.$
electron or hole),  $b_j$ and $f_{j,\alpha}$ are the bosonic and
fermionic partons respectively, $\alpha = \ua, \da$ is the spin
index, and $j$ labels the sites of the triangular lattice. Such a
$Z_2$ spin liquid corresponds to a mean field state of $f_\alpha$
with the following mean field Hamiltonian: $H_{\mathrm{MF}} =
\sum_{\langle i,j \rangle} - t_{ij} f^\dagger_{i,\alpha}
f_{j,\alpha} + \Delta_{ij} (\epsilon_{\alpha\beta} f_{i,\alpha}
f_{j,\beta}) + h.c.$. The parton construction mentioned above
grants $f_{\alpha}$ a local $\U(1)_g$ degrees of freedom, but the
nonzero pairing amplitude $\Delta_{ij}$ breaks the $\U(1)_g$ gauge
symmetry down to $Z_2$, hence the spin liquid phase becomes a
$Z_2$ spin liquid. The bosonic parton $b$ carries both the
physical $\U(1)_e$ charge, and also the $Z_2$ gauge charge. The
SIT in this case corresponds to condensing $b$. The condensation
of $b$ would suppress the $Z_2$ gauge field due to the Higgs
mechanism, and identify $c_\alpha$ and $f_\alpha$. The mean field
Hamiltonian of $f_\alpha$ then becomes the BdG Hamiltonian of the
physical fermion $c_{\alpha}$~\cite{groversl}. It was also shown
numerically that a $Z_2$ spin liquid is indeed in proximity with a
superconductor~\cite{Jiangsl}. Recently this mechanism of SIT has
been explored in the current context of twisted bilayer TMD
moir\'{e} system with a three-orbital model~\cite{kim2024theory}.
\begin{center}
\begin{figure}
\includegraphics[width=0.45\textwidth]{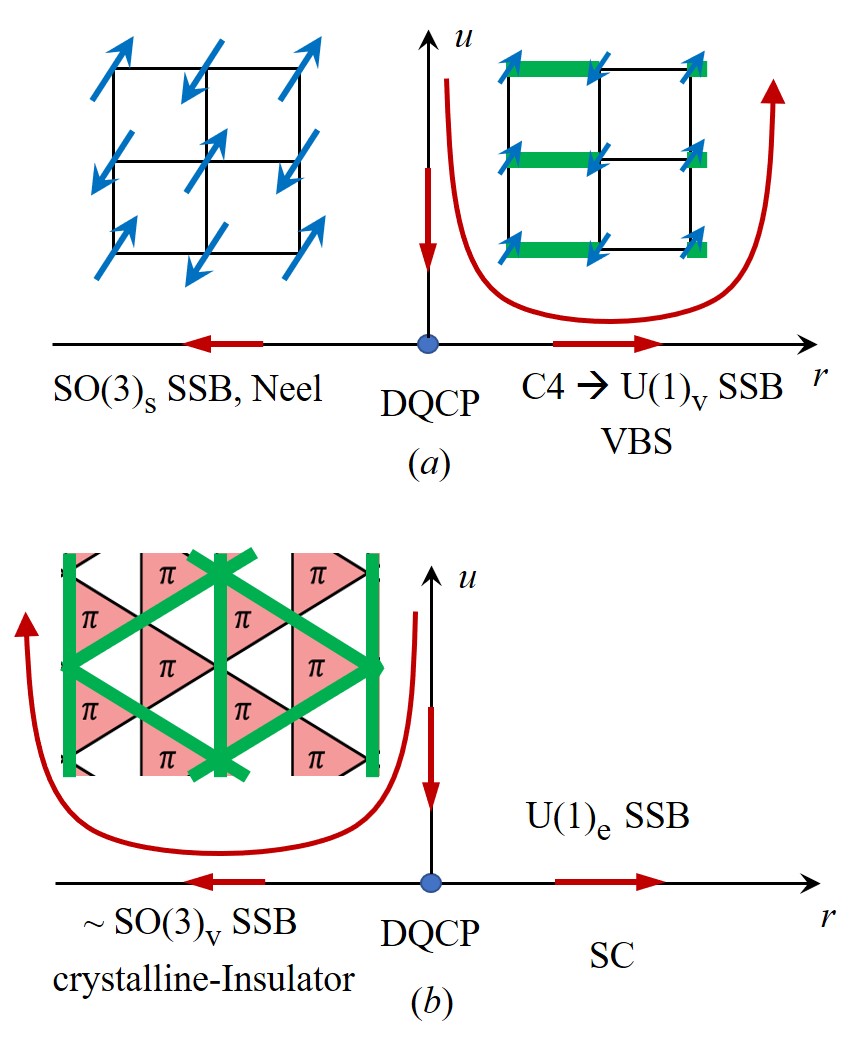}
\caption{Comparison between the original DQCP $(a)$ and the DQCP
proposed in this work $(b)$. In the original DQCP, the transition
is between an AF N\'{e}el order which has spontaneous symmetry
breaking (SSB) of the SO(3)$_s$ spin symmetry, and a VBS order
that spontaneously breaks the $C_4$ rotation symmetry of the
square lattice. Here, the $C_4$ rotation symmetry enlarges into an
emergent $\U(1)_v$ symmetry near the DQCP. In our current case,
the counterpart of the N\'{e}el order is a crystalline order on
the triangular lattice, which is expected to enjoy an approximate
$\SO(3)_v$ symmetry and the role of the $\U(1)_v$ symmetry is
played by the ordinary charge $\U(1)_e$ symmetry. The vertical
axis in $(a)$ represents the term that breaks the $\U(1)_v$
symmetry down to $C_4$, which is dangerously irrelevant as was
discussed in Ref.~\onlinecite{deconfine1,deconfine2}. In our
current case, when $\cL_3$ in Eq.~\ref{L3} is ignored, the
crystalline orders on the triangular lattice will have an emergent
$\SO(3)_v$ symmetry at the SIT. The vertical axis of $(b)$
corresponds to Eq.~\ref{L4} which breaks $\SO(3)_v$ and is
dangerously irrelevant at the DQCP. The operator $\cL_3$ is a
weakly relevant perturbation at the DQCP.} \label{phasedia}
\end{figure}
\end{center}

In this work we propose a new candidate theory for the observed
SIT. We will demonstrate that the observed SIT could actually be a
deconfined quantum critical point (DQCP). In our current system,
though there is no SO(3)$_s$ spin symmetry, there is an
approximate $\SO(3)_v$ symmetry that connects different
crystalline orders, which plays the role as the $\SO(3)_s$
symmetry of the original DQCP. The $\U(1)_v$ symmetry is replaced
by the charge $\U(1)_e$ symmetry (Fig.~\ref{phasedia}). We will
also briefly discuss a second new candidate theory for the SIT,
which is described by an $N_f = 2$ QED, and it is dual to the
easy-plane DQCP. The theory-II applies when the supercondcutor
phase coexists with certain commensurate density wave.

Our proposal is partly motivated by the fact that experimentally
the insulating phase is ``competing" with the superconductor in
the sense that when the SC is suppressed by (for example)
temperature, the insulating phase emerges~\cite{TMDSC}. And the
DQCP is precisely a theory describing the transition between two
competing orders. The experimental signatures of our proposal will
be discussed in more detail in section IV.

\section{A new candidate theory for the SIT: the DQCP}


As was mentioned in the last section, for the purpose of
understanding the universality of the SIT, we model the SIT as
hard-core bosons, or (pseudo) spin-1/2 degrees of freedom on the
effective triangular lattice. The symmetry allowed effective
Hamiltonian of the pseudo spin-1/2 degrees of freedom takes the
following form \beqn H_{\text{eff}} = \sum_{\langle i,j \rangle}
J_z \tilde{S}^z_i \tilde{S}^z_j + J_{xy} (\tilde{S}^x_i
\tilde{S}^x_j + \tilde{S}^y_i \tilde{S}^y_j ) + \cdots \eeqn Here
$\tilde{S}^z_i = n - 1/2$, where $n$ is the density of Cooper
pair; and $\tilde{S}^x_i$, $\tilde{S}^y_i$ are the real and
imaginary parts of the Cooper pair operator. Note that there is a
global constraint that $\sum_i \tilde{S}^z_i = 0$, due to the
fixed half-filling of the charge density. Then one can introduce
the fermionic partons using the standard formalism~\cite{wensl}:
\beqn \tilde{S}^a_i = \frac{1}{2} f^\dagger_{i,\alpha}
\sigma^a_{\alpha\beta} f_{i,\beta}, \eeqn and construct states of
Cooper pair $\tilde{S}^a_i$ using states of $f_\alpha$. This
parton formalism actually has a SU(2)$_g$ invariance, which
becomes explicit when one performs a particle-hole transformation
of $f_\da$: \beqn \psi_{i,1} = f_{i,\ua}, \ \ \ \psi_{i,2} =
f^\dagger_{i,\da}. \eeqn Then the states of $\psi_{i,\alpha}$
could have a maximal SU(2)$_g$ gauge invariance~\cite{wensl}.

The effective (pseudo) spin-1/2 model corresponds to the low
energy bosonic sector of interacting fermions $c_{j,\alpha}$ on
the same triangular lattice. Hence alternatively, one can also
formulate the parton theory directly starting with $c_{j,\alpha}$,
which is a formalism quite broadly used in the studies of cuprates
high temperature
superconductor~\cite{SU2parton,sachdev2009exotic,SU2parton2,SU2parton3,SU2parton4,SU2parton5,SU2parton6,SU2parton7,SU2parton9},
and other exotic quantum liquid
states~\cite{Xumajorana,xuhoneycomb,SU2parton8}: \beqn
c_{j,\alpha} = Z_{j,\alpha\beta} \psi_{j,\beta}, \label{parton1}
\eeqn where $j$ again labels the sites of the effective triangular
moir\'{e} lattice. $Z_{\alpha\beta}$ is a $2\times 2$ SU(2) matrix
field, whose left and right transformations correspond to the spin
symmetry, and a SU(2)$_g$ gauge transformation: \beqn && \SU(2)_s:
Z \ra U_s Z, \cr\cr && \SU(2)_g: Z \ra Z U^{\dagger}_g, \ \ \psi
\ra U_g \psi, \eeqn where $U_s$ and $U_g$ are both SU(2) matrices.
Here we stress that it is {not necessary} for the system to have a
full spin SU(2)$_s$ symmetry, as we are going to consider a phase
diagram where $Z_{\alpha\beta}$ is gapped. Indeed, the TMD
moir\'{e} system has no spin SU(2)$_s$ symmetry due to the
spin-orbit coupling. In addition to the spin and gauge
transformations, $\psi_\alpha$ also carries the charge $\U(1)_e$
symmetry: \beqn \U(1)_e: c_\alpha \ra e^{\ii \theta} c_\alpha, \ \
\ \psi_\alpha \ra e^{\ii \theta} \psi_\alpha. \eeqn

The analysis of a parton formalism always starts with the mean
field state of the partons. In this formalism, the most general
mean field Hamiltonian for $\psi_{j,\alpha}$ takes the following
form \beqn H_{\mathrm{MF}} = \sum_{\langle i,j \rangle}-
t_{ij,\alpha\beta} \psi^\dagger_{i,\alpha} \psi_{j,\beta} + h.c.
\label{MF} \eeqn where \beqn t_{ij,\alpha\beta} \sim \langle
Z^\dagger_{i,\alpha \sigma} t^c_{ij,\sigma\sigma'}
Z_{j,\sigma'\beta} \rangle, \eeqn and $t^c_{ij,\sigma \sigma'}$ is
the hopping amplitude of $c_{j,\alpha}$ on the effective
triangular lattice, which may depend on the spin index $\sigma$
and $\sigma'$.

The goal of this work is not to analyze which mean field state has
the most favorable energy, but to investigate the state that can
potentially lead to an interesting nontrivial SIT. We consider a
mean field state with $t_{ij,\alpha\beta} = \pm t
\delta_{\alpha\beta}$, and the sign $\pm 1$ distribution on the
triangular lattice leads to a staggered $\pi$-flux state for the
mean field band structure of $\psi_\alpha$ (Fig.~\ref{lattice}).
This mean field state preserves the whole $\SU(2)_g$ gauge
symmetry. Here we note that, the staggered $\pi-$flux state
discussed here is different from the standard $\pi-$flux spin
liquid on the triangular lattice often studied in the
literature~\cite{luz2,monopole1,monopole2}, which has $\U(1)_g$
rather than $\SU(2)_g$ gauge invariance. In fact, the state
discussed here would be ``undesirable" in the context of spin
liquid, as this state would break the $\SU(2)_s$ spin symmetry of
the effective spin-1/2 operators $\tilde{S}^a_i$. But again, in
our current system there is no $\SU(2)_s$ symmetry to begin with,
hence the staggered $\pi-$flux state discussed here is legitimate.

It is known that a $\pi$-flux state on the triangular lattice has
two Dirac points in the Brillouin zone, for each component $\alpha
= 1, 2$. The index $\alpha$ is a ``color" index which couples to a
SU(2)$_g$ gauge field. Hence at low energy, the physics of the
mean field state discussed above is expected to be described by
the following Lagrangian: \beqn \cL_{\mathrm{QCD}} = \sum_{v =
1,2} \bar{\psi}_{\alpha,v} \gamma_\mu (\partial_\mu
\delta_{\alpha\beta} - \sum_{l = 1}^3 \ii a^l_\mu
\tau^l_{\alpha\beta} ) \psi_{\beta, v} + \cdots, \label{qcd} \eeqn
where $v = 1,2$ labels the two Dirac valleys. We choose convention
for the gamma matrices, $\gamma_\mu = (-\sigma^2, \sigma^3,
\sigma^1)$ where $\vec{\sigma}$ are Pauli matrices that carry only
Dirac spinor indices (as opposed to $\vec{\tau}$ which are labeled
by color indices).
We note that the two Dirac-valleys discussed here are \textit{not}
the standard valleys associated with the moir\'e superlattice, but
are merely labels for two degenerate Dirac cones in the spectrum
of the staggered $\pi$-flux state.

\begin{figure}
\includegraphics[width=0.43\textwidth]{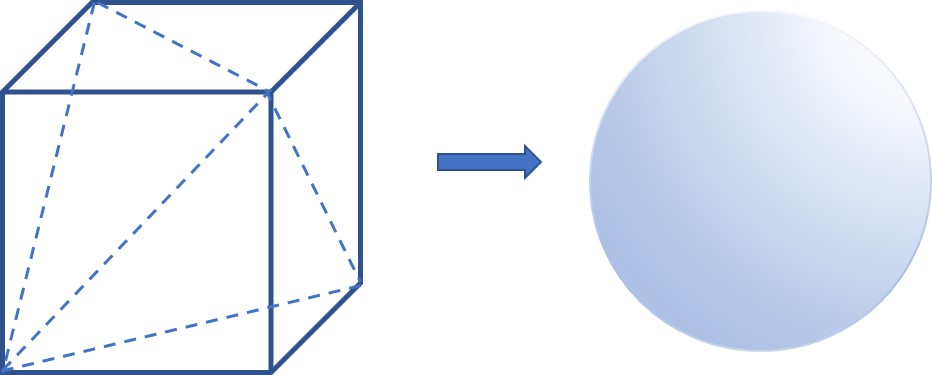}
\caption{The approximate $\SO(3)_v$ symmetry. Due to the existence
of the symmetry allowed term Eq.~\ref{L4}, the three component
vector $\hat{\vect{V}}$ has eight favored degenerate directions
(for $u > 0$), corresponding to the corners of a cube. When
$\cL_3$ is absent, the discrete symmetry of $\hat{\vect{V}}$ will
enlarge into an emergent SO(3)$_v$ at the DQCP, analogous to the
$C_4$ symmetry of the VBS order on the square lattice becomes an
emergent $\U(1)_v$ symmetry at the original DQCP. With the
presence of $\cL_3$, the vector $\hat{\vect{V}}$ prefers to point
towards four corners of a tetrahedron.} \label{emerge}
\end{figure}

The theory in Eq.~\ref{qcd} is $N_f = 2$ QCD with an SU(2)$_g$
gauge field, where the ellipses in Eq.~\ref{qcd} include other
terms allowed by the symmetry, such as the Maxwell term of the
SU(2)$_g$ gauge field, and gauge invariant four-fermion or higher
order terms. As we will show later, one particular combination of
the four-fermion terms will be important, as it is the tuning
parameter of the SIT.

Eq.~\ref{qcd} is the same field theory as the renowned $\pi$-flux
spin liquid state of the spin-1/2 frustrated quantum magnet on the
square lattice. However, we stress that here $\psi$ carries the
charge $\U(1)_e$ symmetry, but no spin quantum number. As was
discussed in detail in Ref.~\onlinecite{SO5}, Eq.~\ref{qcd} is one
of many possible representations of the deconfined quantum
critical point (DQCP), which was originally proposed as a
potential direct continuous quantum phase transition between the
antiferromagnetic N\'{e}el order and the Valence bond solid (VBS)
order of a spin-1/2 frustrated quantum magnet on the square
lattice~\cite{deconfine1,deconfine2}. In the original DQCP, the
key symmetries include a SO(3)$_s$ spin symmetry, and a $C_4$
rotation symmetry of the VBS order, which will be enlarged into a
$\U(1)_v$ symmetry near the DQCP. Furthermore, it was proposed
theoretically~\cite{senthilfisher} and also observed numerically
that~\cite{loopmodel1,loopmodel2}, the $\SO(3)_s \times \U(1)_v$
symmetry further enlarge into an emergent SO(5) symmetry in the
infrared at the DQCP.

In fact, the SO(5) symmetry is the maximal possible continuous
global symmetry of Eq.~\ref{qcd}, which is most clearly revealed
in the Majorana fermion basis. When we write the Dirac fermion
$\psi$ as $\psi = \chi_1 + \ii \chi_2$ with Majorana fermions
$\chi_{1,2}$, another two-component flavor index was introduced,
and there are in total eight components of Majorana fermions
including the gauged color space, and the valley space. The
maximal continuous transformation on the eight Majorana fermions
is SO(8), which includes the SU(2)$_g$ gauge symmetry as its
subgroup. Then, the maximal global symmetry should be a subgroup
of SO(8) which commutes with SU(2)$_g$, and this maximal global
symmetry is SO(5). This symmetry analysis was discussed in detail
in Ref.~\cite{ran2006continuous,xu2008}.

At first glance, it seems rather {\it unlikely} that the current
system under study can realize the DQCP, as the microscopic
symmetry of the TMD moir\'{e} system is too low. For example,
there is no spin SO(3)$_s$ symmetry to begin with. However, in the
following we will demonstrate that, what plays the role of the
$\SO(3)_s$ symmetry of the originally proposed DQCP, is actually
an approximate $\SO(3)_v$ symmetry that transforms among different
crystalline orders on the triangular lattice; and the $\U(1)_v$
symmetry of the original DQCP is replaced by the charge $\U(1)_e$
symmetry.

In fact, it was observed long back that, the crystalline orders on
the triangular lattice may very well enjoy a far larger emergent
symmetry near various critical points or critical
phases~\cite{sondhiz2gauge,xuz2gauge,xutriangle,stiefel}, compared
with the microscopic lattice symmetry. In order to demonstrate
this approximate $\SO(3)_v$ symmetry in our current set-up, let us
consider the following three-component vector of crystalline
order: \beqn \hat{\vect{V}} \sim \bar{\psi}_v\vec{\mu}_{vv'}
\psi_{v'}. \eeqn $\vec{\mu}$ are the three Pauli matrices that
operate on the Dirac-valley space. With a displacement field, the
current moir\'{e} system has the translation symmetries,
time-reversal symmetry, as well as a $C_3$ rotation symmetry of
the triangular lattice. Under these symmetries, the vector
$\hat{\vect{V}}$ transforms as~\footnote{The symmetry of the
effective triangular moir\'{e} lattice is smaller than the full
symmetry of a standard triangular lattice. For the symmetries
under consideration here one can use the projective symmetry group
(PSG) derived for the standard $\pi-$flux spin liquid
state~\cite{luz2,monopole1,monopole2}.} \beqn T_1 &:& (\hV_1,
\hV_2, \hV_3) \ra (- \hV_1, - \hV_2, \hV_3), \cr\cr T_2 &:&
(\hV_1, \hV_2, \hV_3) \ra (- \hV_1, \hV_2, - \hV_3), \cr\cr C_3
&:& (\hV_1, \hV_2, \hV_3) \ra (- \hV_3, \hV_1, - \hV_2), \cr\cr
\cT &:& (\hV_1, \hV_2, \hV_3) \ra (\hV_1, \hV_2, \hV_3).
\label{crystal} \eeqn
Here, $\hV_i$ are density waves at the three $M$ points of the
moir\'{e} Brillouin zone: $K_1 = (\pi/\sqrt{3}, -\pi)$, $K_2 =
(\pi/\sqrt{3}, \pi)$, $K_3 = (2\pi/\sqrt{3}, 0)$
(Fig.~\ref{phasedia}$(b)$). Though these symmetries are smaller
than the full symmetry of a triangular lattice, they already
guarantee that at the quadratic level of $\hat{\vect{V}}$, the
only symmetry allowed term is $|\hat{\vect{V}}|^2$, which has a
$\SO(3)_v$ symmetry, and it should be part of the Lagrangian in
Eq.~\ref{qcd}. Higher order terms of $\hat{\vect{V}}$ may break
the $\SO(3)_v$ symmetry, but they are at least sixth order of
$\psi$. For example the following term  \beqn \cL_4 \sim u
(|\hV_1|^4 + |\hV_2|^4 + |\hV_3|^4) \label{L4} \eeqn is allowed by
the microscopic symmetry, but it is a product of eight fermions,
and is expected to be highly irrelevant for the $N_f = 2$ QCD
Eq.~\ref{qcd}. Another term which is the product of six fermions
is also allowed by symmetry: \beqn \cL_3 \sim u' \hV_1 \hV_2
\hV_3. \label{L3}\eeqn We will discuss the role of $\cL_3$ at the
DQCP later.

\begin{figure}

\includegraphics[width=0.42\textwidth]{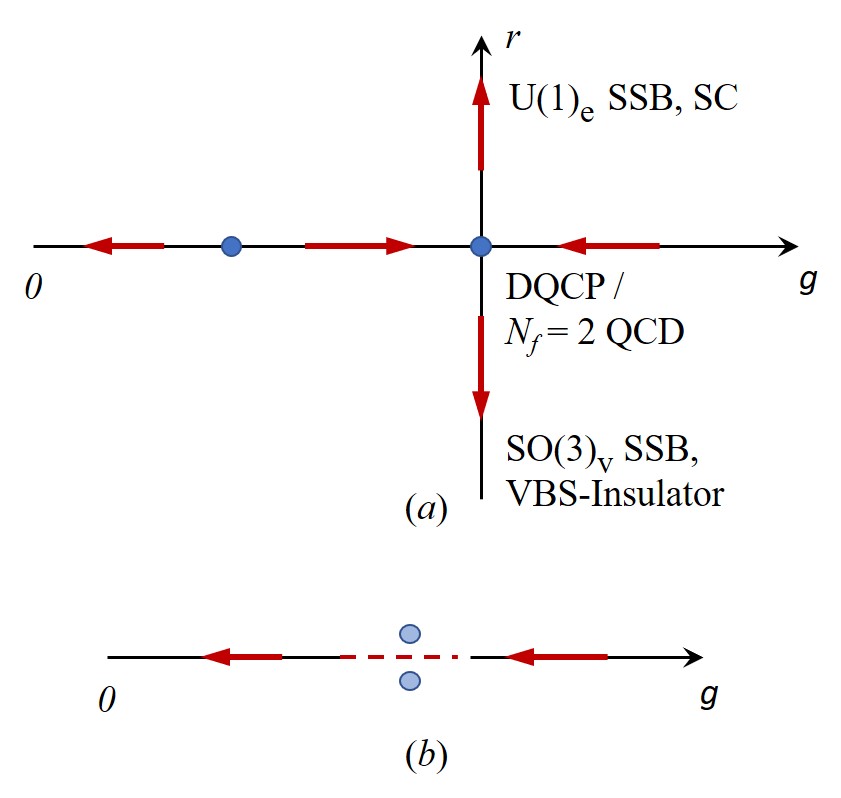}

\caption{($a$). The originally expected RG flow of $g$ in
Eq.~\ref{NLSM}. There is a full SO(5) symmetry along the axis $r =
0$, and there are two fixed points while increasing $g$ from zero,
the one with larger $g$ corresponds to the DQCP.
The tuning parameter $r$ in Eq.~\ref{deltaL} tunes the transition
between the superconductor and the crystalline insulator. ($b$). A
possible scenario of the RG flow based on up-to-date
understanding. The two fixed points in $(a)$ annihilate and become
a pair of nearby complex fixed points, causing $g$ to flow very
slowly in the dashed region. This ``walking" RG flow leads to an
approximate conformal symmetry. } \label{RG}

\end{figure}

The $\U(1)_v$ symmetry of the original DQCP now becomes the charge
$\U(1)_e$ symmetry, and the order parameter of the $\U(1)_e$
symmetry are the real and imaginary parts of the inter-valley
gauge singlet $s-$wave superconductor Cooper pair: \beqn
\hat{\Delta} \sim \epsilon_{\alpha\beta} \psi^t_{\alpha v}
\gamma_0 \mu^2_{vv'} \psi_{\beta v'}. \eeqn The crystalline
symmetries in Eq.~\ref{crystal}, the $\U(1)_e$ symmetry, and
time-reversal together guarantee that all fermion-bilinear terms
are not allowed in Eq.~\ref{qcd}, except for one: the chemical
potential for charge density $\mu \bar{\psi}\gamma_0\psi$. But the
chemical potential is tuned to zero in the system, due to the
fixed half-filling of charge density.

To make further connections with DQCP, we define a five-component
unit vector $\vect{n} = (n_1, \cdots n_5)$ with $|\vect{n}|^2 =
1$, and couple it to the fermion bilinear operators as \beqn \cL
&=& \cL_{\mathrm{QCD}} + \frac{1}{g} (\partial_\mu \vect{n})^2
\cr\cr &+& m \bigg  (\sum_{a = 1}^3 n_a \hV_a + n_4
\mathrm{Re}[\hat{\Delta}] + n_5 \mathrm{Im}[\hat{\Delta}] \bigg).
\eeqn Following the procedure of Ref.~\onlinecite{abanov2000},
integrating out the fermions would lead to the following low
energy effective nonlinear sigma model (NLSM) field theory for
$\vect{n}$ with a Wess-Zumino-Witten term at level-1: \beqn
\cL_{\mathrm{eff}} &=& \frac{1}{g} (\partial_\mu \vect{n})^2
\cr\cr &+& \frac{2\pi \ii}{\Omega_4} \int_0^1 du \epsilon_{abcde}
n^a \partial_x n^b \partial_y n^c \partial_\tau n^d \partial_u
n^e. \label{NLSM} \eeqn The NLSM above is another ``classic"
description of the DQCP which well captures the intertwinement of
the order parameters involved in the DQCP~\cite{senthilfisher}.

The relation between the NLSM, the $N_f = 2$ QCD, and the DQCP is
shown in the proposed RG flow diagram of $g$ of Eq.~\ref{NLSM}.
The originally proposed RG flow of $g$ is that
(Fig.~\ref{RG}($a$)), if one starts with small $g$ where the NLSM
is in its ordered phase, increasing $g$ is expected to first drive
an order-disorder phase transition; and due to the existence of
the WZW term in Eq.~\ref{NLSM}, the disordered phase was expected
to be controlled by a stable fixed point with SO(5) symmetry,
which corresponds to the $N_f = 2$ QCD and is also the DQCP tuned
to the critical point. The tuning parameter of the DQCP, which
tunes the N\'{e}el-VBS transition in the original DQCP, and the
SIT transition in our current context, corresponds to the
following term: \beqn \delta \cL &\sim& r (n_1^2 + n_2^2 +n_3^2 -
n_4^2 - n_5^2) \cr\cr &\sim& r (|\hat{\vect{V}}|^2 -
|\hat{\Delta}|^2) = r (|\bar{\psi} \vec{\mu}\psi|^2 - |\psi^t
\gamma_0 \mu^2 \epsilon \psi|^2). \label{deltaL} \eeqn The first
and second line of Eq.~\ref{deltaL} are supposed to be the tuning
parameters of the DQCP in Eq.~\ref{NLSM} and Eq.~\ref{qcd}
respectively.

To investigate the details of the crystalline order ($r < 0$ in
Eq.~\ref{deltaL}), we need to analyze the role of the higher order
term Eq.~\ref{L4} in the crystalline ordered phase. This term is
irrelevant at the DQCP, but it will play important role in the
phase with $r < 0$, i.e. $u$ is actually ``dangerously
irrelevant". For $r < 0$ and $u > 0$, the system would have an
eight fold degenerate crystalline order, corresponding to \beqn
\langle \hat{\vect{V}} \rangle &\sim& V (\pm 1, \pm 1, \pm 1);
\eeqn while for $u < 0$, there are six degenerate ground states
with \beqn \langle \hat{\vect{V}} \rangle &\sim& V (\pm 1, 0, 0),
\ V (0, \pm 1, 0), \ V (0, 0, \pm 1). \eeqn A nonzero $\cL_3$ in
Eq.~\ref{L3} would lift the degeneracy, and favor $\vect{\hV}$ to
point along four diagonal directions of a cube, which form a
tetrahedron (Fig.~\ref{emerge}).

Here, we would also like to clarify the nature of the DQCP, which
has been an extensively studied subject in the last two decades.
As we explained before, the originally expected RG flow of the
coupling constant $g$ of the NLSM is shown in
Fig.~\ref{RG}($a$)~\cite{Ma_2020,Nahum_2020}, with two fixed
points one of which is the DQCP and also the $N_f = 2$ QCD. But
since the early days of the numerical simulation of the
N\'{e}el-VBS DQCP, which is supposedly realized in the $J-Q$ model
on the square lattice~\cite{JQ1,JQ2}, people have noticed that
though many aspects of the transition appear to be continuous,
there are always various difficulties fitting the data with the
standard $(2+1)d$ CFT. For example, the critical exponents are
drifting with the system
size~\cite{JQ3,JQ4,loopmodel1,loopmodel2}, and the exponents
extracted seem incompatible with the bounds established by the
conformal Bootstrap~\cite{bootstrap,bootstrap1,libootstrap}. Also,
nonlocal probes such as entanglement entropy at the DQCP are found
to be inconsistent with a $(2+1)d$
CFT~\cite{Zhao_2022,song2024extracting,song2024deconfined,dengee,demidio2024entanglement}.
In Fig.~\ref{RG}($b$) we sketch one of the possible scenarios of
the RG flow of $g$ based on the up-to-date understanding: the two
fixed points in Fig.~\ref{RG}$(a)$ in fact annihilate and become a
pair of nearby complex fixed points, causing $g$ to flow very
slowly in the dashed region~\cite{SO5}. This slow RG flow due to
the annihilation of real fixed points, and emergence of complex
fixed points is referred to as the ``loss of
conformality"~\cite{conformallost}. But with a large and finite
system size, we are still expected to see critical behavior due to
the slow RG flow of $g$, even though eventually the DQCP is an
extremely weak first order transition~\cite{1storder,sandvik1st}.
In fact, the approximate conformal symmetry has been well-observed
in the fuzzy sphere realization~\cite{hefuzzy} of Eq.~\ref{NLSM},
even though there is a weakly relevant SO(5) singlet. We also note
that, recently another possible scenario for DQCP was proposed,
that there exists a nearby multi-critical point with a full SO(5)
symmetry~\cite{chester2024,sandvik1st}.

Now let us discuss the role of $\cL_3$ in Eq.~\ref{L3} at the
DQCP. At first glance, this term is a product of six fermions,
which also appears highly irrelevant. However, a more careful
investigation shows that, assuming the DQCP has an emergent SO(5)
symmetry, then $\cL_3$ belongs to the symmetric rank-3 tensor
representation of the SO(5) which is in the same representation of
the triple-monopole in the CP$^1$ formalism of the DQCP and the
triple-monopole is allowed by symmetry for the N\'{e}el-VBS DQCP
realized on a honeycomb lattice. Both the Monte-Carlo study and
the fuzzy sphere realization of the DQCP suggest that the
triple-monopole of the CP$^1$ representation of the DQCP is weakly
relevant~\cite{monoscaling,hefuzzy,chen2024emergent}. For example,
the Monte Carlo determines that the triple-monopole has scaling
dimension 2.80(3)~\cite{monoscaling}. Hence $\cL_3$ should also be
a weakly relevant perturbation at the DQCP. Nevertheless, previous
numerical works suggest that even with this weakly relevant
perturbation, the DQCP may still behave like a continuous
transition over a large window of length scale. For example,
signatures of DQCP were also found at the N\'{e}el-VBS transition
on the honeycomb lattice~\cite{dqcphoneycomb1,dqcphoneycomb2}.

\section{Signatures of the DQCP}

In this section we discuss the experimental signatures of the
proposed candidate theory for SIT. In particular, we would like to
distinguish the current theory of SIT from the previously known
theory of SIT between the superconductor and a $Z_2$ spin liquid
state discussed in the introduction.

The most obvious difference between these two theories is that,
the SC-to-$Z_2$ spin liquid theory should be insensitive to the
lattice symmetry breaking. In particular, if the $Z_2$ spin liquid
and the superconductor are both gapped, the nature of the
transition should belong to the $3D$ XY$^\ast$ transition, whose
universality class does not change when the lattice symmetry is
weakly broken. But our current DQCP theory is more sensitive to
the lattice symmetry. For example, let us consider an external
strain field that breaks the $C_3$ rotation symmetry. This strain
field acts like a term $ - s |V_3|^2$ in the Lagrangian, which
favors $V_3$ over $V_1$ and $V_2$ if $s > 0$ and the DQCP now has
an ``easy-axis" anisotropy. In the NLSM description of the DQCP,
this term is $ - s (n_3)^2$, and is a strongly relevant
perturbation.

\begin{figure}
\includegraphics[width=0.4\textwidth]{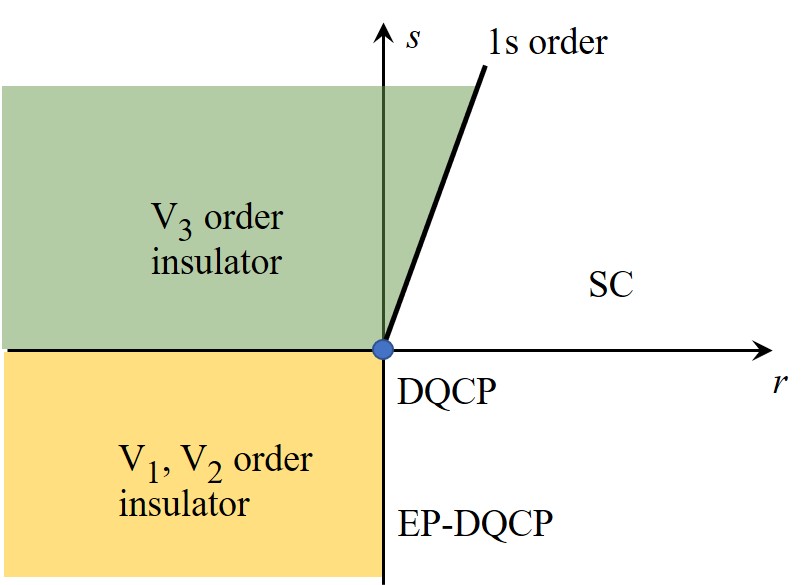}
\caption{The phase diagram around the DQCP under strain. The
parameter $s$ in Eq.~\ref{dLp} is tuned by the strain. A positive
$s$ will shift the transition point, and drive the DQCP into a
first order transition between the SC and an insulator with the
$V_3$ order; a negative $s$ will drive the DQCP into the
easy-plane DQCP between the SC and an insulator with the $V_1,
V_2$ order. } \label{phasedia4}
\end{figure}

A relevant perturbation on a critical point can lead to several
possible consequences. At the DQCP, leveraging the SO(5) emergent
symmetry, one can conclude that, with a positive $s$, the DQCP
becomes a first order transition between the SC and an insulator
with the order of $V_3$, and the transition point is shifted from
$r = 0$. This analysis was given in Ref.~\onlinecite{SO5} in the
context of the original DQCP on the square lattice. In the current
case, let us turn on both the tuning parameter $r$ of the DQCP,
and the extra strain which breaks the $C_3$ rotation symmetry:
\beqn \delta \cL' = - s (n_3)^2 - r (n_4^2 + n_5^2). \label{dLp}
\eeqn Along the axis $s = r > 0$, there is an emergent SO(3)
symmetry that rotates among $\vec{n} = (n_3, n_4, n_5)$, and the
system would develop an order of $\vec{n}$ which spontaneously
breaks the emergent SO(3). Now the line $r = s$ becomes the
transition line, and tuning $r$ away from $s$ would drive a
prominent first order transition between the order of $(n_4, n_5)$
(the superconductor) and the order of $n_3$, i.e. an insulator
with the crystalline order of $\langle V_3 \rangle \neq 0$. Hence
with the strain, one should see hysteresis when scanning through
the SIT.

Let us also consider a strain along the opposite direction, i.e.
adding a term $ - s (n_3)^2$ with negative $s$ in the Lagrangian.
As was discussed thoroughly in the literature of the DQCP, this
term would drive the DQCP to an easy-plane DQCP. The easy-plane
DQCP is still a direct transition between the SC and the
crystalline insulator with $V_1$ or $V_2$ order. Again, though the
easy-plane DQCP should still be a very weak first order
transition~\cite{EPnumeric,zhao2024scaling,nahumO4,fakherO4}, it
may appear continuous for a large length scale. And very
nontrivial physics can still occur at the easy-plane DQCP,
including an emergent O(4)
symmetry~\cite{EPnumeric,O4,fakherO4,nahumO4,dqcpex2}. Although $
\cL_3$ in Eq.~\ref{L3} is a weakly relevant perturbation at the
full DQCP, to the best knowledge of the authors, it is unclear
whether $\cL_3$ is relevant or irrelevant at the easy-plane DQCP.
The predicted phase diagram under strain in sketched in
Fig.~\ref{phasedia4}.

Another difference between the DQCP and the SC-to-Z2 spin liquid
transition theory is that, our current theory of DQCP is a theory
of ``competing orders". Indeed, the experiments suggest that, when
the SC is suppressed, the insulating state emerges. In fact, in
our theory the SC and the crystalline phases not only compete with
each other but also have a mutual 't Hooft anomaly, in the sense
that the defect of one order parameter should carry the quantum
number of the other. For example, the vortex of the SC carries the
fractionalized quantum number of $\hat{\vect{V}}$, which should
have nonzero lattice momentum. Hence if one creates a vortex
inside the superconductor, the vortex core should have crystalline
order.

\section{The second candidate theory}

In this section we propose a second candidate theory for the observed SIT, but the nature of the SC phase and the crystalline-insulator are different from the previous theory. 
To facilitate the discussion, we take a different parton
construction from Eq.~\ref{parton1}: \beqn c_{j,\ua} = b_j
\psi_{j,1}, \ \ \ c_{j,\da} = b_j^\dagger \psi_{j,2}.
\label{parton2} \eeqn We still consider mean field state of
$\psi_\alpha$ with the same form of mean field Hamiltonian as
Eq.~\ref{MF}, with $t_{ij,\alpha\beta} = \pm t
\delta_{\alpha\beta}$. The sign of $t_{ij,\alpha\beta}$ still
constitutes a staggered $\pi-$flux shown in
Fig.~\ref{lattice}($b$), except now the gauge field fluctuation is
$\U(1)_g$ rather than $\SU(2)_g$. The low energy physics of this
state is given by $N_f = 4$ QED, where the U(1) gauge field
couples via $\tau^3$ in Eq.~\ref{qcd}. One can also obtain the
same theory by condensing a Higgs field in Eq.~\ref{qcd} that
breaks the $\SU(2)_g$ gauge group down to $\U(1)_g$.

Now, we consider the adding the following extra fermion bilinear
terms to $N_f = 4$ QED: \beqn \delta \cL = m_1 \bar{\psi} \tau^3
\psi + m_2 \bar{\psi} \mu^3 \psi. \eeqn The first term $m_1
\bar{\psi} \tau^3 \psi$ is actually allowed by the symmetry and
gauge invariance, and hence it should exist on top of the $N_f =
4$ QED.  The second term $m_2 \bar{\psi} \mu^3 \psi$ is precisely
$\hV_3$, and it breaks part of the lattice symmetry as it carries
finite momentum.

We assume that $m_1$ and $m_2$ are both {\it finite} and {\it
positive}, and we consider the tuning parameter $\delta m = m_1 -
m_2$. Near the critical point $\delta m = 0$, of the four Dirac
fermions, the two with quantum numbers $\tau^3 \mu^3 = +1$ are
well-gapped, while the other two with $\tau^3\mu^3 = -1$ have low
energy and mass $\delta m$. Hence, the low energy theory
describing this transition is $N_f = 2$ QED:
\beqn \cL_{\mathrm{QED}} = \bar{\psi}_\alpha \gamma_\mu
(\partial_\mu \delta_{\alpha\beta} - \ii a_\mu
\tau^3_{\alpha\beta} ) \psi_\beta + \delta m
(\bar{\psi}\tau^3\psi) + \cdots \label{qed}\eeqn Eq.~\ref{qed} has
``half" of the fermion modes of Eq.~\ref{qcd}: the fermion in
Eq.~\ref{qed} with $\tau^3 = +1$ corresponds to the Dirac fermion
in Eq.~\ref{qcd} with $\tau^3 = +1$, $\mu^3 = -1$; and the fermion
in Eq.~\ref{qed} with $\tau^3 = -1$ correspond to the Dirac
fermion with $\tau^3 = -1$, $\mu^3 = +1$ in Eq.~\ref{qcd}.

The two phases in the phase diagram with positive and negative
$\delta m$ are identified as the SC phase and a
crystalline-insulator phase respectively. When $\delta m > 0$,
i.e. $m_1>m_2>0$, the gauge flux of $a_\mu$ will have the charge
of $\psi$. More precisely, it carries charge-$2e$ under the
$U(1)_e$ symmetry due to the dual quantum spin Hall effect (i.e.,
the spin-flux carries electric charge). The gauge flux of $a_\mu$
is generically a conserved quantity, and the photon phase of
$a_\mu$ is dual to the condensate of gauge fluxes. This condensate
is equivalently a charge-$2e$ superconductor due to the dual
quantum spin Hall effect. Hence we identify the phase with $\delta
m > 0$ as a superconductor phase. In fact, the superconductor
phase arises here for essentially the same reason as Skyrmion
condensation induced superconductivity, when the Skyrmion carries
electric
charges~\cite{groversenthil,Khalaf_2021,qshscfakher,qshscfakher2}.
It was also shown before that a ``gauged" quantum spin Hall state
can lead to inplane N\'{e}el order and
superconductor~\cite{ransu2}.

The critical point of Eq.~\ref{qed} is located at $\delta m = 0$,
which is $N_f = 2$ QED. It was shown that $N_f = 2$ QED is
self-dual~\cite{xudual,seiberg2,mrossdualPRL}, and hence it was
conjectured to possess an O(4) emergent symmetry in the infrared.
This emergent O(4) symmetry was seen
numerically~\cite{EPnumeric,O4,fakherO4,nahumO4}. It was also
shown that the $N_f = 2$ QED is dual to the easy-plane Abelian
Higgs model with two flavors of scalar fields (also referred to as
the easy-plane noncompact CP$^1$ model)~\cite{potterdual,SO5},
which describes the DQCP with an easy-plane anisotropy.

\begin{figure}
\includegraphics[width=0.4\textwidth]{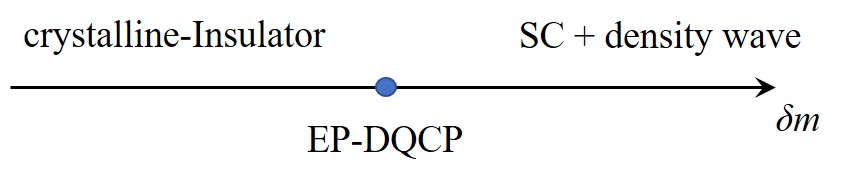}
\caption{The phase diagram of the second candidate theory of the
SIT, which corresponds the $N_f = 2$ QED, and it is also dual to
the easy-plane DQCP.} \label{phasedia2}
\end{figure}

To identify more details of the SC phase, we need to analyze the
monopole quantum number of the $\U(1)$ gauge field $a_\mu$, with
$\delta m > 0$. Most importantly, we would like to know whether
the Cooper pair condenses at zero momentum, or finite momentum.
This not an easy task, but results from previous studies that
thoroughly analyzed the monopole quantum numbers of the staggered
$\pi$-flux Dirac spin liquid on the triangular lattice may be
used. It was shown that~\cite{monopole1,monopole2}, with a quantum
spin Hall mass term, $m_1 \bar{\psi} \tau^3 \psi$, the $\pi-$flux
state would have nonzero staggered charge density
$\bar{\psi}\gamma_0 \psi$ at the center of each triangle of the
lattice. This charge distribution leads to finite momentum carried
by the monopole operator of the gauge field considered therein.
However, in our current case, we argue that since the gauge field
is generated by $\tau^3$, the extra charge density
$\bar{\psi}\gamma_0 \psi$ should be ``invisible" to the gauge
fluxes. We conjecture here that the superconductivity induced by
the dual-QSH effect due to the QSH mass term $m_1$ and the gauge
fluctuation of $a_\mu \tau^3$ would still order at zero momentum.

Of course, we assumed that there is a nonzero $m_2$ throughout the
phase diagram, and a nonzero $m_2$ leads to a density wave order
in the superconductor. Hence in this second theory, the
superconductor has coexisting Cooper pair condensate and a density
wave order at momentum $K_3 = (2\pi/\sqrt{3}, 0)$.

When $\delta m < 0$, i.e. $m_1 < m_2$, there is no dual quantum
spin Hall effect and the superconductor disappears. However, the
monopole of the $\U(1)$ gauge field will still carry crystalline
symmetry quantum number due to the fermion modes trapped by the
monopole. In this case, there is further crystalline symmetry
breaking due to the flux condensate of the $\U(1)$ gauge field.
Further analysis analogous to
Ref.~\onlinecite{monopole1,monopole2} is demanded to fully
determine the crystalline order in this case.

\section{Summary and discussion}

In this work we propose that the superconductor-to-insulator
transition observed recently in the twisted bilayer TMD moir\'{e}
system may be described by the DQCP, which is a subject that has
attracted enormous attention and efforts in multiple disciplines
of physics in the last two decades. We propose that the
crystalline order on the effective triangular lattice has an
approximate $\SO(3)_v$ symmetry, which is the counterpart of the
$\SO(3)_s$ spin symmetry of the original DQCP; and the charge
$\U(1)_e$ symmetry plays the role as the emergent $\U(1)_v$
symmetry that rotates between VBS orders in the original DQCP. We
also demonstrated that the DQCP may be driven into either a
prominent first order transition, or an easy-plane DQCP under
strain.

Physics discussed in this work may also have applications in other
systems, especially when the system can be modelled by strongly
interacting bosons on the triangular lattice at half-filling. We
note here that our theory is different from the previously
discussed superfluid-insulator transition for bosons on the
triangular lattice~\cite{burkovbalents}. Potential connections
between these two formalisms will be explored in the future.

We also briefly discussed a second candidate theory for the SIT,
where the superconductor is induced by a ``dual" quantum spin Hall
effect, meaning the spin gauge field fluctuation would lead to
charge condensate, i.e. a superconductor. To fully identify the
nature of the superconductivity, and the crystalline order of the
insulator phase of the second theory, we need detailed analysis of
the monopole quantum numbers of the gauge field, which we defer to
future study.

The authors thank Matthew Fisher, Chao-Ming Jian, Xue-Yang Song
for very helpful discussions. C.X. acknowledges support from the
Simons foundation through the Simons investigator program.

\bibliography{SIT}

\end{document}